\documentclass[aps,preprintnumbers,showpacs,twocolumn,amsmath,nofootinbib,amssymb,floatfix]{revtex4}
\usepackage{graphicx,color,dcolumn,booktabs,bm}
\usepackage{epstopdf}
\usepackage{longtable,lscape}
\usepackage{txfonts}
\usepackage{overpic}
\usepackage{float}
\usepackage{amssymb}
\usepackage{amsmath}
\usepackage{indentfirst}
\usepackage{feynmf}  
\usepackage{slashed}  
\usepackage{cases}
\usepackage{ulem}
\usepackage{color}
\usepackage{float}
\usepackage{multirow}
\usepackage{tikz}
\usepackage{epsfig,dsfont,amssymb,amsmath,amsfonts,amsbsy,mathrsfs}
\usepackage{multirow}
\usepackage{graphicx}  
\usepackage{float}  
\usepackage{subfigure}  
\usepackage{array}
\usepackage{microtype}
\usepackage{pdfpages} 
\usepackage{flushend}
\usepackage[colorlinks, citecolor=blue,anchorcolor=red,menucolor=red, linkcolor=red,filecolor=red,runcolor=red,urlcolor=blue,frenchlinks=true]{hyperref}

\allowdisplaybreaks[4]

\begin{document}

\title{Threshold effects as the origin of $Y(4500)$ observed in $e^+e^-\to J/\psi K^+K^-$}

\author{Xiao-Yun Wang$^{1}$}
\author{Liu-Lin Wang$^{1}$}
\author{Xiao-Hai Liu$^{1}$}~\email{xiaohai.liu@tju.edu.cn}
\author{Qiang Zhao$^{2,3,4}$}\email{zhaoq@ihep.ac.cn}

\affiliation{ $^{1}$Center for Joint Quantum Studies and Department of Physics, School of Science, Tianjin University, Tianjin 300350, China\\
$^{2}$ Institute of High Energy Physics and Theoretical Physics Center for Science Facilities,
Chinese Academy of Sciences, Beijing 100049, China\\
$^{3}$ University of Chinese Academy of Sciences, Beijing 100049,  China\\
$^{4}$ Center for High Energy Physics, Henan Academy of Sciences, Zhengzhou 450046,  China
}

\date{\today}
\begin{abstract}

The BESIII collaboration has recently observed a resonant structure $Y(4500)$ in $e^{+}e^{-}\to J/\psi K^{+}K^{-}$, whose origin remains unresolved. In this study, we analyze the cross section line shape of $e^{+}e^{-}\to J/\psi K^{+}K^{-}$ by taking into account the $\psi(4415)$ state and intermediate charmed meson loops. By treating $\psi(4415)$ as both a pure $S$-wave state and $S$-$D$ mixed states, and introducing the $Z_{cs}^{(\prime)}$ resonance, we find that the $Y(4500)$ peak at $\sqrt{s}=4.5~\text{GeV}$ arises from the triangle singularity mechanism in the $Z_{cs}^{(\prime)}$-mediated rescattering processes, supporting its interpretation as a threshold effect rather than a conventional resonance.

\end{abstract}

\maketitle

\section{Introduction}
\label{sec:introduction}

Over the past two decades, numerous states incompatible with the conventional quark model—referred to as the $XYZ$ states—have been discovered experimentally. These observations have stimulated considerable interest both theoretically and experimentally~\cite{Chen:2016qju,Chen:2016spr,Esposito:2016noz,Guo:2017jvc,Olsen:2017bmm,Brambilla:2019esw,Chen:2022asf}. Most of these $XYZ$ states exhibit properties inconsistent with conventional quark model configurations, sparking intense debate regarding their internal structures. The fundamental question of whether these states are genuine exotic hadrons—such as tetraquarks, molecular states, or hybrids—or merely kinematic threshold effects arising from rescattering processes remains unresolved.

Many $XYZ$ states lie near hadron-pair thresholds, prompting their interpretation as hadronic molecules. Alternatively, some non-resonance interpretations were also proposed due to the nontrivial near-threshold phenomena, such as the threshold cusp or triangle singularity (TS) mechanism. In particular, the TS mechanism can produce resonance-like peaks in invariant mass spectra, complicating the identification of exotic states~\cite{Liu:2019dqc,Liu:2020orv,Liu:2016xly,Wu:2011yx,Wang:2013cya,Liu:2014spa,Liu:2013vfa,Guo:2014iya,Ketzer:2015tqa,Szczepaniak:2015eza,Guo:2015umn,Liu:2015taa,Liu:2015fea,Achasov:2015uua,Guo:2016bkl,Aceti:2012dj,Aceti:2016yeb,Bayar:2016ftu,Roca:2017bvy,Liu:2016dli,Liu:2015cah,Cao:2017lui,Dong:2020hxe,Nakamura:2021qvy}. It is therefore essential to evaluate such possibilities before attributing a peak to a true particle. We refer to Ref.~\cite{Guo:2019twa} for a comprehensive review of the threshold cusp and TS in hadronic reactions.

The TS mechanism in multi-body rescattering processes are capable of mimicking exotic hadron signals without physical resonances. They occur in loop diagrams when three intermediate particles can simultaneously approach on-shell under specific kinematic conditions—the so-called ``two-cut'' condition~\cite{Eden:1966dnq}. This results in nonanalytic behavior of the scattering amplitude, manifesting as sharp cusps or threshold enhancements in mass spectra. For instance, in $Y(4260) \to J/\psi\pi\pi$ decays, TS effects near the $\bar{D}D^*$ threshold produce a peak resembling the $Z_c(3900)$, despite the absence of a bound state \cite{Liu:2013vfa,Liu:2014spa,Wang:2013cya}. The TS mechanism has been applied across various systems, both in the light hadron spectroscopy and heavy flavor sector. The effects induced by the TS depend sensitively on kinematics, coupling strengths, and the widths of intermediate states.

Among the enigmatic states, the $Y$ family with $J^{PC}=1^{--}$ produced directly in $e^+ e^-$ collisions holds a special position. Their production mechanisms and decay modes offer a valuable laboratory for investigating the nature of exotic hadrons~\cite{Guo:2025ady}. The well-known $Y(4260)$ state (also referred to as $Y(4230)$ in later measurements) was first observed by the BaBar collaboration in the $e^+ e^- \to J/\psi \pi^+ \pi^-$ process~\cite{BaBar:2005hhc}, and its existence was subsequently confirmed by the CLEO~\cite{CLEO:2006tct}, Belle~\cite{Belle:2007dxy,Belle:2013yex}, and BESIII~\cite{BESIII:2016bnd} collaborations. The unusual properties of the $Y(4260)$ have stimulated extensive theoretical studies (see e.g. recent reviews in Refs.~\cite{Brambilla:2019esw,Wang:2025clb}). Recent observations by the BESIII collaboration have further enriched this field. In 2022, using 15.6 fb$^{-1}$ of data, BESIII measured the $e^{+}e^{-}\to J/\psi K^{+}K^{-}$ cross sections at center-of-mass energies from 4.127 to 4.600 GeV. Two resonant structures were observed: one consistent with the known $Y(4230)$, and a new structure, denoted as $Y(4500)$, with a statistical significance exceeding $8\sigma$. Its mass and width were determined to be $(4484.7 \pm 2.3 \pm 21.5)$ MeV and $(111.1 \pm 30.1 \pm 15.2)$ MeV, respectively~\cite{BESIII:2022joj}. Further investigations revealed additional structures: the $Y(4710)$ was discovered in the $e^{+}e^{-}\to J/\psi K^{+}K^{-}$ process within the center-of-mass energy range of 4.61 to 4.95 GeV~\cite{BESIII:2023wqy}, while the $\psi(4470)$ resonance was observed in $e^+ e^- \to D^{*0} D^{*-} {\pi}^+$ with mass $4469.1 \pm 26.2 \pm 3.6$ MeV and width $246.3 \pm 36.7 \pm 3.6$ MeV~\cite{BESIII:2023cmv}. Although the mass of $\psi(4470)$ is compatible with that of $Y(4500)$, its width is significantly larger, and its decay rate to $D^* {\bar D}^* \pi$ is two orders of magnitude greater than the decay rate of $Y(4500)$ to $J/\psi K^+ K^- $. Additionally, a resonance with mass consistent with $Y(4500)$ has been observed in $e^+ e^- \to { J/\psi \pi}^+ {\pi}^-$, though with only $3\sigma$ statistical significance~\cite{BESIII:2022qal}.
The emergence of these structures revives the long-standing question: Are they genuine exotic hadrons, conventional charmonia, or merely threshold enhancements caused by rescattering effects?

The nature of the newly observed $Y(4500)$ state has been explored through multiple theoretical frameworks. In Ref.~\cite{Deng:2023mza}, $Y(4500)$ was interpreted as a $\psi_1(3D)$ charmonium state using an unquenched quark model that incorporates coupled-channel effects, with predictions given for its mass and width. A similar charmonium picture is presented in Ref.~\cite{Wang:2022jxj}, where $Y(4500)$ is described as a $5S$–$4D$ mixed state with a fitted mass of $4504 \pm 4$ MeV and a predicted width of 36–45 MeV.
Beyond conventional charmonia, tetraquark interpretations have also been proposed. Ref.~\cite{Chiu:2005ey} estimated the mass of a $cs\bar{c}\bar{s}$ state to be about $4450 \pm 100$ MeV, while Ref.~\cite{Wang:2024qqa} used QCD sum rules to suggest that $Y(4500)$ may be a $c\bar{c}q\bar{q}$ tetraquark with a width around 200 MeV.
Additionally, a molecular picture has been put forward in several studies~\cite{Peng:2022nrj,Gungor:2023ksu,Wang:2016wwe,Dong:2021juy}, in which $Y(4500)$ is viewed as a $D_{s1}(2536)D_s$ molecular state.

The $e^+e^- \to J/\psi K^+K^-$ process is also particularly interesting as one of the promising channels for discovering exotic states such as the $Z_{cs}$ state. The $Z_{cs}(3985)$ was first observed by the BESIII Collaboration in 2020~\cite{Ablikim:2020hsk} in the $K^+$ recoil-mass spectra of the process $e^+e^- \to K^+(D_s^- D^{*0} + D_s^{*-}D^0)$. Its mass and width were measured to be $(3982.5^{+1.8}_{-2.6}\pm 2.1)$ MeV and $(12.8^{+5.3}_{-4.4}\pm 3.0)$ MeV, respectively. Various theoretical interpretations have been proposed for the nature of the $Z_{cs}(3985)$, including the $D_s D^*$/$D_s^*D$ molecular picture~\cite{Meng:2020ihj,Yang:2020nrt,Du:2020vwb,Chen:2020yvq,Sun:2020hjw,Guo:2020vmu,Yan:2021tcp}, the tetraquark state hypothesis~\cite{Wan:2020oxt,Wang:2020rcx,Wang:2020iqt,Jin:2020yjn}, threshold effects~\cite{Yang:2020nrt,Ikeno:2021ptx}, reflection effects~\cite{Wang:2020kej}, among others. Another state, the $Z_{cs}(4000)^+$ with quark content $c\bar{c}u\bar{s}$, was reported by the LHCb collaboration in the decay $B^+ \to J/\psi \phi K^+$~\cite{Aaij:2021ivw}. Its mass and width are $(4003\pm 6^{+4}_{-14})$ MeV and $(131\pm 15\pm 26)$ MeV, respectively. Although the mass of the $Z_{cs}(4000)$ is close to that of the $Z_{cs}(3985)$, it has a significantly larger width. Whether the $Z_{cs}(3985)$ and $Z_{cs}(4000)$ are the same state remains an open question~\cite{Chen:2022asf,Du:2022jjv,Yang:2020nrt,Ortega:2021enc,Meng:2021rdg,Ikeno:2021mcb}. In Ref.~\cite{Ortega:2021enc}, the authors employed a coupled-channel model, and their results suggest that the $Z_{cs}(3985)$ and $Z_{cs}(4000)$ are likely the same state. The Particle Data Group (PDG) has classified both under the label $T_{c\bar{c}\bar{s}1}(4000)$, reporting average values of mass and width as $M = 3988 \pm 5~\text{MeV}$ and $\Gamma = 14 \pm 4~\text{MeV}$, respectively~\cite{ParticleDataGroup:2024cfk}.
Within the framework of SU(3) flavor symmetry, the $Z_c(4020)$ state, located near the $D^* D^*$ threshold, is expected to have a strange partner ($J^P=1^+$) denoted as $Z_{cs}^{\prime}$, lying near the $D^* D_s^*$ threshold. Although this state has not yet been experimentally observed, its properties have been explored in theoretical studies~\cite{Ortega:2021enc,Yang:2020nrt,Ozdem:2021hka}. In our analysis, we estimate the mass of $Z_{cs}^{\prime}$ by applying the same mass shift relative to the corresponding threshold as observed for the $Z_{cs}$ state. For the width, we assume it to be comparable to that of $Z_{cs}$.

In this work, we analyze the $e^{+}e^{-}\to J/\psi K^{+}K^{-}$ reaction via the rescattering processes, and try to figure out whether the threshold effects can induce the $Y(4500)$ signal. The impact of $Z_{cs}^{(\prime)}$ state in this reaction is also studied.

This paper is organized as follows. In Sec.~\ref{sec:framework}, we introduce the theoretical framework used in the analysis. Numerical results are presented in Sec.~\ref{sec:results}, and a summary is provided in Sec.~\ref{sec:summary}.

\section{Theoretical Framework}
\label{sec:framework}

\begin{figure*}[htbp]
	\includegraphics[width=0.6\textwidth]{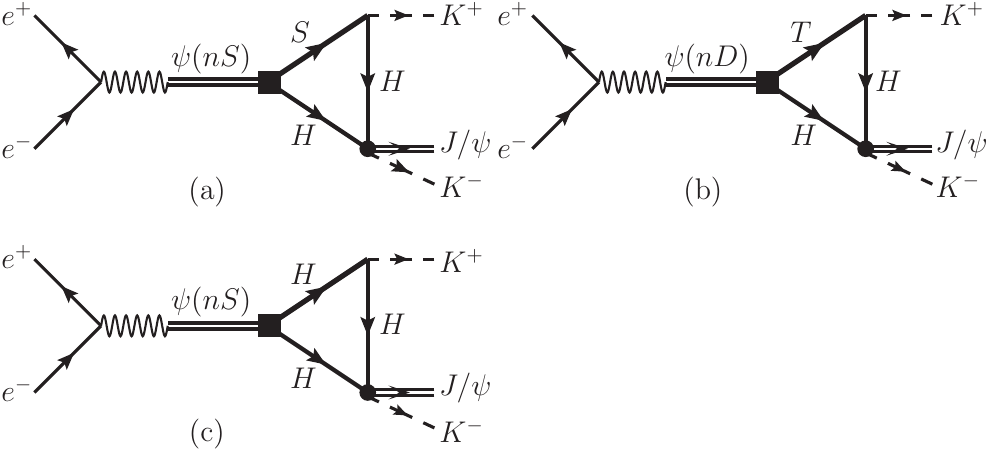}
	\caption{ Feynmann diagrams for $e^{+}e^{-}\to J/\psi K^{+}K^{-}$ via the intermediate charmonium state $\psi(nS/nD)$ and charmed-meson loops. The charge conjugate diagrams are implicit.}
	\label{fig:e+e-}
\end{figure*}

For $e^{+}e^{-}\to J/\psi K^{+}K^{-}$, we take into account contributions from various charmed meson loops, as illustrated in Fig.~\ref{fig:e+e-}. This reaction proceeds via the experimentally well-established charmonium state, such as the $\psi(4415)$, whose mass is close to the $Y(4500)$ resonance of interest. In our model, the $\psi(4415)$ is treated as a 
conventional charmonium state, which could be an S-wave state $\psi(nS)$, a D-wave state $\psi(nD)$, or a mixture.

To calculate the rescattering processes in Fig.~\ref{fig:e+e-}, our theoretical framework is based on the effective field theory with the heavy quark spin symmetry (HQSS) involved.

In the study of heavy-light mesons $Q \bar q$ under HQSS, the heavy quark can be treated as a spectator in the heavy quark limit $M_Q \to \infty$. In this limit, the heavy quark's influence is simplified, and the interactions are dominated by the light quark. At this point, the interaction between the heavy quark and gluons decouples, and the heavy quark spin $s_Q$ becomes a conserved quantity. Consequently, the total angular momentum $J$ of the hadron can be written as
\begin{eqnarray}
    J &=& s_Q + j_q, 
\end{eqnarray}   
where $j_q = L + s_q$ is the total angular momentum of the light degrees of freedom, $s_Q$ is the heavy quark spin, and $s_q$ is the light quark spin. Therefore, information about hadrons can be described by the spin and parity quantum numbers of the light degrees of freedom. 

The ground-state heavy mesons comprise the $J^P = 0^-$ and $J^P = 1^-$ states, which form a doublet under the classification $j_q^{P} = \frac{1}{2}^-$. In heavy hadron chiral perturbation theory (HHChPT), these states are incorporated into a superfield defined as~\cite{Casalbuoni:1996pg,Wise:1992hn,Burdman:1992gh,Yan:1992gz} 
\begin{equation}
H_{1a} = \frac{1 + \slashed{v}}{2} \left[ \mathcal{D}_{a\mu}^* \gamma^\mu - \mathcal{D}_a \gamma_5 \right],
\end{equation}
where $a$ is the light-flavor index labeling the light quark component, and $v$ is the heavy quark velocity. Here, $\mathcal{D}_a$ and $\mathcal{D}_{a\mu}^*$ represent the pseudoscalar and vector charmed mesons, respectively. Explicitly, $\mathcal{D}_a$ corresponds to the multiplet $(D^0, D^+, D_s^+)$, while $\mathcal{D}_{a\mu}^*$ denotes $(D^{*0}, D^{*+}, D_s^{*+})$. The charge conjugate superfield is given by
\begin{equation}
H_{2a} = \left[ \bar{\mathcal{D}}_{a\mu}^* \gamma^\mu + \bar{\mathcal{D}}_a \gamma_5 \right] \frac{1 - \slashed{v}}{2}.
\end{equation}

The doublet with light degrees of freedom $j_q^P={\frac{1}{2}}^+$ is composed of mesons with $J^P$=$0^+$ and $J^P$=$1^+$. The superfield and the corresponding charge conjugate field read
\begin{eqnarray}
S_{1a} &=& \frac{1+\slashed{v}}{2} [{\cal D}_{1a}^{\prime \mu} \gamma_{\mu} {\gamma}_5 - {\cal D}_{0a}],\\
S_{2a} &=& [{\bar{\cal D}}_{0a}-{\bar{\cal D}}_{1a}^{\prime\mu} \gamma_\mu \gamma_5 ] \frac{1-\slashed{v}}{2}.
\end{eqnarray}
The doublet with the light degrees of freedom $j_q^P={\frac{3}{2}}^+$ consists of mesons whose quantum numbers are $J^P$=$1^+$ and $J^P$=$2^+$. These states are combined into the superfields 
\begin{eqnarray}
T_{1a}^{\mu} &=&  \frac{1+\slashed{v}}{2} \big[{\cal D}_{2a}^{\mu \nu} {\gamma}_{\nu} - \sqrt{\frac{3}{2}} {\cal D}_{1a\nu} {\gamma}_5 [g^{\mu \nu} - \frac{1}{3} {\gamma}^{\nu} ({\gamma}^{\mu} - v^{\mu})] \big],\nonumber \\
T_{2a}^{\mu} &=&  [ {\bar{\cal D}}_{2a}^{\mu \nu} {\gamma}_{\nu} + \sqrt{\frac{3}{2}} {\bar{\cal D}}_{1a\nu} {\gamma}_5 [g^{\mu \nu} - \frac{1}{3} {\gamma}^{\nu} ({\gamma}^{\mu} - v^{\mu})]] \frac{1-\slashed{v}}{2}.\nonumber \\
\end{eqnarray}
For simplicity, in this work we use $HH$ to represent a combination of $D_{(s)}^{(*)} D_{(s)}^{(*)}$ meson pair. The symbols $SH$ and $TH$ have similar conventions. 

The HQSS is also applicable to heavy quarkonia. The charmonia states $\psi(nS)$ and $\psi(nD)$ are grouped into the multiplet fields $J$ and $J^{\mu\nu}$, respectively, which take the following forms~\cite{DeFazio:2008xq}
\begin{eqnarray}
J &=& \frac{1+\slashed{v}}{2} [\psi(nS)^\mu \gamma_\mu - \eta_c(nS) \gamma_5] \frac{1-\slashed{v}}{2},\\
J^{\mu \nu} &=& \frac{1+\slashed{v}}{2} \Bigg\{\psi(nD)_\alpha \Bigg[ \frac{1}{2} \sqrt{\frac{3}{5}}[(\gamma^\mu - v^\mu) g^{\alpha\nu} + (\gamma^\nu \nonumber\\
&& - v^\nu) g^{\alpha\mu} - \sqrt{\frac{1}{15}} (g^{\mu\nu} - v^{\mu} v^{\nu}) \gamma^{\alpha} \Bigg] \Bigg\} \frac{1-\slashed{v}}{2}.
\end{eqnarray}
Only the expression relevant to $\psi(nD)$ in $J^{\mu\nu}$ is presented here.

\subsection{Effective couplings}

The leading-order effective Lagrangian for the coupling between $S$-wave and $D$-wave charmonia and the charmed mesons is given by
\begin{eqnarray}
&&{\cal L}_{\psi}  = \frac{g_{TH}}{\sqrt{2}} \langle J^{\mu \nu} {\bar H}_{2a} {\gamma}_{\nu} {\bar T}_{1a\mu} - J^{\mu \nu} {\bar T}_{2a\mu} {\gamma}_{\nu} {\bar H}_{1a} \rangle  \nonumber\\
&&+ C_S \langle J {\bar H}_{2b} \gamma_\mu \gamma_5 {\bar H}_{1a} {\cal A}^\mu_{ba} \rangle + i g_{HH} \langle J^{\mu \nu} {\bar H}_{2a} \gamma_\mu {\mathop{\partial}\limits^{\leftrightarrow}}_\nu {\bar H}_{1a} \rangle  \nonumber\\
&&+ g_{SH} [\langle J {\bar S}_{2a} {\bar H}_{1a} + J {\bar H}_{2a} {\bar S}_{1a} \rangle]
+ \mathrm{H.c.}, \label{eq:psi}
\end{eqnarray}
with
\begin{eqnarray}
&&{\bar H}_{1a, 2a}= {\gamma}^0 H_{1a,2a}^{\dagger} {\gamma}^0, \\
&&{\bar S}_{1a, 2a}= {\gamma}^0 S_{1a,2a}^{\dagger} {\gamma}^0, \\
&&{\bar T}_{1a, 2a}^\mu= {\gamma}^0 T_{1a,2a}^{\dagger\mu} {\gamma}^0,
\end{eqnarray}
where $\langle...\rangle$ means the trace on the Dirac matrix. 
In Eq.~(\ref{eq:psi}), the term $C_S \langle J {\bar H}_{2b} \gamma_\mu \gamma_5 {\bar H}_{1a} {\cal A}^\mu_{ba} \rangle$ describes the $HHJ/\psi K$ contact interaction in Fig.~\ref{fig:e+e-}, where
${\cal A}^\mu$ is the chiral axial vector containing the Goldstone bosons
\begin{eqnarray}
{\cal A}_\mu = \frac{1}{2}(\xi^\dagger \partial_\mu \xi - \xi \partial_\mu \xi^\dagger),
\end{eqnarray}
with
\begin{eqnarray}
&&\xi = e^{i{\cal M}/f_{\pi}},\\
&&{\cal M} = \left( \begin{array}{ccc} \frac{1}{\sqrt{2}}\pi^0 + \frac{1}{\sqrt{6}}\eta & \pi^+ &K^+ \\\pi^- & -\frac{1}{\sqrt{2}}\pi^0 + \frac{1}{\sqrt{6}}\eta & K^0 \\K^- & {\bar K}^0 & -\sqrt{\frac{2}{3}}\eta \end{array} \right).
\end{eqnarray}

According to HQSS, both the heavy quark spin and the total angular momentum are conserved, implying that the total angular momentum of the light degrees of freedom must also be conserved. The $S$-wave charmonium state $\psi(nS)$ can couple to $SH$ via $S$-wave interaction while preserving HQSS. Such contributions are taken into account, as depicted in Fig.~\ref{fig:e+e-}(a). In contrast, the coupling between $\psi(nS)$ and $TH$ must occur in relative $D$-wave to satisfy HQSS constraints. Since higher partial waves are generally suppressed near threshold, we neglect loop diagrams involving $\psi(nS)TH$ vertices. On the other hand, the $D$-wave charmonium $\psi(nD)$ may couple to $TH$ in relative $S$-wave while preserving HQSS. Therefore, contributions from $\psi(nD)TH$ loop diagrams are included in our calculation, as shown in Fig.~\ref{fig:e+e-}(b). The $\psi(nS)$ can couple to $HH$ via $P$-wave interaction, and the corresponding contributions are illustrated in Fig.~\ref{fig:e+e-}(c).

The explicit loop diagrams corresponding to Fig.~\ref{fig:e+e-} are presented in Table~\ref{tab:diagram}. It should be noted that contributions from certain loop diagrams—such as $D_0 D^* [D_s]$, $D_1 D [D_s^*]$, and others in Figs.~\ref{fig:e+e-}(a) and (b)—were found to be numerically small. Similarly, the $HHH$-loop contribution from Fig.~\ref{fig:e+e-}(c) is also negligible. These contributions are therefore omitted from the final results. This is because the TSs of these diagrams are far from the physical region in the complex energy plane.

\begin{table}
    \centering
    \caption{Explicit Feynman diagrams included in the calculation for Fig.~\ref{fig:e+e-}. Each diagram is labeled by a charmed meson loop. The particle inside the bracket corresponds to the vertical propagator in the triangle loop.}
    \begin{ruledtabular}
    \begin{tabular}{c|l}
        Diagram & Triangle loop \\ 
        \hline  
        Fig.~\ref{fig:e+e-}(a) & $D_{s0}^* D_s^* [D]$, $D_{s1}^{\prime} D_s [D^*]$, $D_{s1}^{\prime} D_s^* [D^*]$ \\
        Fig.~\ref{fig:e+e-}(b) & $D_{s1} D_s [D^*]$, $D_{s1} D_s^* [D^*]$, $D_{s2} D_s^* [D]$, $D_{s2} D_s^* [D^*]$  \\
        Fig.~\ref{fig:e+e-}(c) & $D_{s} D_s [D^*]$, $D_{s} D_s^* [D^*]$, $D_{s}^* D_s^* [D]$, $D_{s}^* D_s^* [D^*]$
    \end{tabular}
    \end{ruledtabular}      
    \label{tab:diagram}
\end{table}

\begin{table}
\caption{Effective coupling constants for the $\psi(nS)SH$ and $\psi(nD)TH$ vertices, computed within the $^3P_0$ model. Values are given in GeV.}
\label{tab:coupling}
\begin{ruledtabular}
\begin{tabular}{lcccc}
$\psi$&Channel	&Coupling& Channel&Coupling\\
\colrule
$\psi(4S)$&$D_s^* D_{s0}^*$&$5.31$ & $D_s D_{s1}^{\prime}$ & $1.05$\\
$ $ &$D_s^* D_{s1}^{\prime}$&$1.09$ & $$ & $$\\
$\psi(5S)$&$D_s^* D_{s0}^*$&$2.69$ & $D_s D_{s1}^{\prime}$ & $0.54$\\
$ $ &$D_s^* D_{s1}^{\prime}$&$0.55$ & $ $ & $ $\\
$\psi(3D)$ &$D_s D_{s1}$&$3.26$	&$D_s^* D_{s1}$& $3.38$\\
$ $ &$D_s^* D_{s2}$&$4.19$	&$$& $$\\
$\psi(4D)$ &$D_s D_{s1}$&$1.47$	&$D_s^* D_{s1}$& $1.02$\\
$ $ &$D_s^* D_{s2}$&$1.87$	&$$& $$\\
\end{tabular}
\end{ruledtabular}
\end{table} 
For the coupling constants of the $\psi(nS)SH$ and $\psi(nD)TH$ vertices given in Eq.~(\ref{eq:psi}),  we employ the $^{3}P_{0}$ model to evaluate them. By matching the decay amplitude derived from the $^{3}P_{0}$ model to the effective Lagrangian given in Eq.~(\ref{eq:psi}), we determine the corresponding effective couplings, which are summarized in Table~\ref{tab:coupling}. The detailed computational procedure is provided in Appendix~\ref{3P0 Model}. It should be pointed out that when using the $^{3}P_{0}$ model to estimate these coupling constants, the HQSS is broken to some extent; nevertheless, we retain the relative phase between different couplings as derived from Eq.~(\ref{eq:psi}).

In our numerical calculation, the physical states $D_{s0}^{*}(2317)$ and $D_{s1}(2460)$ are identified with the members of the $j_q^P=\frac{1}{2}^+$ doublet $(D_{s0}^*,D_{s1}^\prime)$, and the $D_{s1}(2536)$ and $D_{s2}^*(2573)$ form the doublet $(D_{s1},D_{s2})$ with $j_q^P=\frac{3}{2}^+$. There are two $J^P = 1^+$ axial-vector states $D_{s1}^\prime$ and $D_{s1}$, which are assumed to be 
coherent superpositions of quark model spin-singlet and spin-triplet states,
\begin{eqnarray}
    |D_{s1}(2460)\rangle &=& -| ^1P_1 \rangle \sin\phi + |^3P_1\rangle \cos\phi,\\
    |D_{s1}(2536)\rangle &=& | ^1P_1 \rangle \cos\phi + |^3P_1\rangle \sin\phi.
\end{eqnarray}
In the heavy quark limit, we have the mixing angle $\phi \simeq 35.3^\circ$~\cite{Godfrey:1986wj}.

The relevant strong interaction terms in HHChPT are given by the Lagrangian
\begin{eqnarray}
{\cal L}_{\mathcal{D}} &=& i\frac{h^\prime}{\Lambda_\chi} \langle {\bar H}_{1a} T_{1b}^{\mu} {\gamma}_\nu {\gamma}_5 (D_\mu {\cal A}_\nu + D_\nu {\cal A}_\mu)_{ba} \rangle \nonumber\\
&& + ig \langle H_{2b} \gamma_\mu \gamma_5 {\cal A}_{ba}^\mu {\bar H}_{1a} \rangle + i h \langle S_{1b} \gamma_\mu \gamma_5 {\cal A}_{ba}^\mu {\bar H}_{1a} \rangle,
\end{eqnarray}
where the coupling constant $h$ is related to the strong decay rate of the charmed meson and can be determined from experimental data. We adopt the average value of $h$ from Ref.~\cite{Mehen:2004uj}, namely $h^2 = 0.44 \pm 0.11$. The coupling $h^\prime/\Lambda_\chi$ is taken to be 0.55 GeV$^{-1}$~\cite{Casalbuoni:1996pg}.

According to the vector meson dominance (VMD) model~\cite{Bauer:1975bw,Li:2007au}, the interaction between a photon and a vector meson is described by the Lagrangian
\begin{eqnarray}
{\cal L}_{V \gamma} &=& \frac{e M_V^2}{f_V} V_{\mu} A^{\mu},
\end{eqnarray}
where $M_V$ and $f_V$ denote the mass and decay constant of the vector meson $V$, respectively. In the limit of vanishing electron mass ($m_e \approx 0$), the decay constant $f_V$ can be extracted from the decay width of $V \to e^+ e^-$
\begin{eqnarray}
\frac{e}{f_V} &=& \left[\frac{3 \Gamma_{V \to e^+ e^-}}{\alpha M_V}\right]^{1/2},
\label{eq:fv}
\end{eqnarray}
with $\alpha = 1/137$ being the fine-structure constant. Using the measured value $\Gamma_{\psi(4415) \to e^+ e^-} = 0.58 \pm 0.07$ keV~\cite{ParticleDataGroup:2024cfk}, we obtain $e/f_{\psi(4415)} = 0.0073\pm 0.0004$.

\begin{figure}[tb]
	\includegraphics[width=0.27\textwidth]{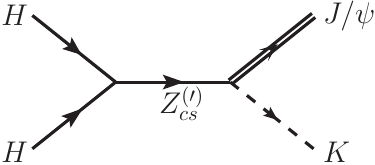}
	\caption{ Diagrammatic representation of the rescattering channel via the state $Z_{cs}^{(\prime)}$, showing the subprocess $HH \to J/\psi K$.}
	\label{fig:zcs}
\end{figure}

Whether the $Z_{cs}^{(\prime)}$ state exists highly affect the rescattering amplitude. If it exists, the $HHJ/\psi K$ vertex in Fig.~\ref{fig:e+e-} can be replaced by the diagram depicted in Fig.~\ref{fig:zcs}, and this kind of pole contribution will be dominant in the near-threshold energy region.

The coupling constants of $Z_{cs}$ with $J/\psi K$ and $D_s^*D$ ($D^* D_s$) channels can be estimated by employing the compositeness relation~\cite{Weinberg:1962hj}.  The total compositeness relation for the two-channel scenario reads~\cite{Meissner:2015mza,Guo:2019kdc}
\begin{eqnarray}
    X= X_1 + X_2 =  |g_1|^2 \Big|\frac{\partial G_1^{II}(s_R)}{\partial s} \Big| +  |g_2|^2 \Big|\frac{\partial G_2^{II}(s_R)}{\partial s} \Big|,
    \label{eq:X}
\end{eqnarray}
where $ G_j^{II}(s)$ represents the one-loop two-point function $G(s)$ on the second Riemann sheet~\cite{Oller:2000fj,Oller:2006jw}. The $g_j$ ($j=$1,2) denotes the effective coupling between the two-particle state and the resonance. The lighter channel is labeled 1, and the heavier channel is labeled 2. In this work, we assume the $Z_{cs}$ is a resonance and follow the method proposed in Refs.~\cite{Guo:2020vmu,Kang:2016ezb} to estimate these couplings. The $X$ can be calculated by the scattering length $a$ and the effective range $r$
\begin{eqnarray}
    X &=& \left( \frac{2 r}{a}-1 \right)^{-1/2}.
\end{eqnarray}
For a resonance pole at $E_R=M_R-i\Gamma_R/2$, the $a$ and $r$ are determined as
\begin{eqnarray}
   && a=-\frac{2k_i}{k_r^2+k_i^2},\ \ r=-\frac{1}{k_i}, \\
   && k_r+ik_i\equiv \sqrt{\frac{2m_1 m_2}{m_1+m_2}(E_R-m_1-m_2)}\ ,
\end{eqnarray}
with $m_1$ and $m_2$ the masses of the scattering states. The saturation condition of the $Z_{cs}$ decay width by the 
$J/\psi K$ and $D_s^*D$ ($D^* D_s$) channels gives
\begin{eqnarray}
    &&\Gamma_R = \Gamma_1 + \Gamma_2 \nonumber \\
    && = |g_1|^2 \frac{q_1(M_R^2)}{8 \pi M_R^2}+ |g_2|^2 \int_{m_{th}}^{M_R + 2 \Gamma_R} dE \frac{q_2 (E^2)}{16 {\pi}^2 E^2} \frac{\Gamma_R}{ (M_R - E)^2 + \frac{\Gamma_R^2}{4}}.\nonumber\\
    \label{eq:Gamma}
    \end{eqnarray}
Combining the above equations allows for solving $g_1$ and $g_2$. The estimation on the coupling constants of $Z_{cs}^\prime$ shares a similar formalism.

\subsection{Rescattering amplitudes}

The leptonic and hadronic parts of Fig.~\ref{fig:e+e-} are connected by a single photon, and  the cross section can be written as the product of leptonic $L^{\mu\nu}(k_1, k_2)$ and hadronic $H_{\mu\nu}(Q)$~\cite{CTEQ:1993hwr}
\begin{eqnarray}
\sigma = L^{\mu\nu}(k_1,k_2) H_{\mu\nu}(Q),
\end{eqnarray}
where $k_1$ and $k_2$ are the leptons' momenta and $Q=k_1+k_2$.
Ignoring the electron mass, the lepton part $L^{\mu\nu}$ can be written as
\begin{eqnarray}
L^{\mu\nu}(k_1, k_2) &=& \mathrm{Tr}[\slashed{k_1} \gamma^\mu \slashed{k_2} \gamma^\nu] \nonumber\\
&=& 4(k_1^\mu k_2^\nu + k_1^\nu k_2^\mu - \frac{s_1}{2} g^{\mu\nu}),
\end{eqnarray}
with $s_1=Q^2$ being the square of the center of mass energy. We introduce the Breit-Wigner (BW) function for the intermediate vector charmonium state $\psi(4415)$ when calculating the scattering amplitude.

The transition amplitudes of $\psi(nS)\to  J/\psi K^+ K^-$ via the $SHH$ loops are given by
\begin{eqnarray}
&&{\cal M}_{D_{s0}^* D_s^* [D]} = -4 i \int \frac{d^4q}{(2\pi)^4} g_{SH} \frac{h C_S}{f_{\pi}^2}\epsilon^{\tau_1}(\psi) \epsilon^{\delta_3}(J/\psi) (v\cdot p_1)\nonumber \\
&&\times (v\cdot p_3) g_{\tau_1 \lambda_1} g_{\delta_3 \lambda_3} \frac{(-g^{\lambda_1 \lambda_3} + v^{\lambda_1} v^{\lambda_3})}{[q_1^2-m_{D_s^*}^2][q_2^2-m_{D_{s0}^*}^2][q_3^2-m_{D}^2]},\\
&&{\cal M}_{D_{s1}^{\prime} D_s [D^*]} =  4 i \int\frac{d^4q}{(2\pi)^4}g_{SH} \frac{h C_S}{f_{\pi}^2} \epsilon^{\tau_1}(\psi) \epsilon^{\delta_3}(J/\psi)(v\cdot p_1) g_{\tau_1 \lambda_1} \nonumber \\
&&\times (v\cdot p_3) g_{\alpha_2 \beta_2} g_{\delta_3 \lambda_3} \frac{(- g^{\beta_2 \lambda_3} + v^{\beta_2} v^{\lambda_3}) (-g^{\lambda_1 \alpha_2} + v^{\lambda_1} v^{\alpha_2})}{[q_1^2-m_{D_s}^2][q_2^2-m_{D_{s1}^{\prime}}^2][q_3^2-m_{D^*}^2]},\\
&&{\cal M}_{D_{s1}^{\prime} D_s^* [D^*]} = -4 i \int\frac{d^4q}{(2\pi)^4}g_{SH} \frac{h C_S}{f_{\pi}^2} \epsilon_{\tau_1}(\psi) \epsilon_{\delta_3}(J/\psi) \epsilon^{\alpha_1 \beta_1 \lambda_1 \tau_1} \nonumber \\
&&\times \epsilon^{\alpha_3 \lambda_3 \beta_3 \delta_3} v_{\alpha_3} g^{\alpha_2 \beta_2}v_{\beta_1} (v\cdot p_1) (v\cdot p_3) (-g_{\alpha_1 \alpha_2} + v_{\alpha_1} v_{\alpha_2}) \nonumber \\
 &&\times \frac{(- g_{\lambda_1 \beta_3} + v_{\lambda_1} v_{\beta_3}) (- g_{\beta_2 \lambda_3} + v_{\beta_2} v_{\lambda_3})}{[q_1^2-m_{D_s^*}^2][q_2^2-m_{D_{s1}^{\prime}}^2][q_3^2-m_{D^*}^2]},
\end{eqnarray}
where the subscript on $\mathcal{{M}}$ indicates the corresponding triangle diagram listed in Table~\ref{tab:diagram}.

The amplitudes of $\psi(nD)\to  J/\psi K^+ K^-$ via the $THH$ loops read
\begin{eqnarray}
&&{\cal M}_{D_{s1} D_s [D^*]} = - 4\sqrt{\frac{10}{3}} \int\frac{d^4q}{(2\pi)^4} g_{TH} \frac{h^{\prime}}{\Lambda_{\chi}} \frac{C_S}{f_{\pi}^2} \epsilon^{\tau_1}(\psi) \epsilon^{\delta_3}(J/\psi) \nonumber \\
&& \times g_{{\nu}_1 {\tau}_1} [3 p_{1 \lambda_2} p_{1 \nu_2} + ((v \cdot p_1)^2 - p_1^2) g_{{\lambda_2}{\nu_2}}] (v \cdot p_1) g_{{\delta_3}{\lambda_3}} \nonumber \\
&& \times \frac{(-g^{\nu_1 \nu_2}+ v^{\nu_1} v^{\nu_2})(-g^{\lambda_2 \lambda_3}+ v^{\lambda_2} v^{\lambda_3})}{[q_1^2-m_{D_s}^2][q_2^2-m_{D_{s1}}^2][q_3^2-m_{D^*}^2]},\\
&&{\cal M}_{D_{s1} D_s^* [D^*]} = -2 \sqrt{\frac{10}{3}} \int\frac{d^4q}{(2\pi)^4} g_{TH} \frac{h^{\prime}}{\Lambda_{\chi}} \frac{C_S}{f_{\pi}^2}  v_{\alpha_1} (v \cdot p_3) \epsilon_{\tau_1}(\psi) \nonumber \\
&& \times \epsilon_{\delta_3}(J/\psi) \epsilon^{\nu_1 \tau_1 \alpha_1 \beta_1} [3 p_1^{\lambda_2} p_1^{\nu_2} + ((v \cdot p_1)^2 - p_1^2) g^{{\lambda_2}{\nu_2}}] \epsilon^{\alpha_3 \lambda_3 \beta_3 \delta_3} v_{\alpha_3} \nonumber \\
&& \times (-g_{\nu_1 \nu_2}+ v_{\nu_1} v_{\nu_2})  \frac{(-g_{\beta_1 \beta_3}+ v_{\beta_1} v_{\beta_3})(-g_{\lambda_2 \lambda_3}+ v_{\lambda_2} v_{\lambda_3})}{[q_1^2-m_{D_s^*}^2][q_2^2-m_{D_{s1}}^2][q_3^2-m_{D^*}^2]},\\
&&{\cal M}_{D_{s2} D_s^* [D]} = -4 \sqrt{\frac{6}{5}} \int\frac{d^4q}{(2\pi)^4} g_{TH} \frac{h^{\prime}}{\Lambda_{\chi}} \frac{C_S}{f_{\pi}^2} \epsilon_{\tau_1}(\psi) \epsilon^{\delta_3}(J/\psi) \nonumber \\
&& \times g^{{\nu}_1 {\tau}_1} g_{\delta_3 \lambda_3} p_1^{\mu_2} p_1^{\nu_2} (v \cdot p_3) \Bigg\{ \frac{1}{2} \Big[ (-g_{\mu_1 \mu_2} + v_{\mu_1} v_{\mu_2}) (-g_{\nu_1 \nu_2} + v_{\nu_1} v_{\nu_2}) \nonumber \\
&& + (-g_{\mu_1 \nu_2} + v_{\mu_1} v_{\nu_2}) (-g_{\nu_1 \mu_2} + v_{\nu_1} v_{\mu_2}) \Big]  - \frac{1}{3} (-g_{\mu_1 \nu_1} + v_{\mu_1} v_{\nu_1})\nonumber \\
&& \times (-g_{\mu_2 \nu_2} + v_{\mu_2} v_{\nu_2}) \Bigg\}  \frac{(-g^{\mu_1 \lambda_3} + v^{\mu_1} v^{\lambda_3})}{[q_1^2-m_{D_s^*}^2][q_2^2-m_{D_{s2}}^2][q_3^2-m_{D}^2]},\\
&&{\cal M}_{D_{s2} D_s^* [D^*]} = -4 \sqrt{\frac{6}{5}} \int\frac{d^4q}{(2\pi)^4} g_{TH} \frac{h^{\prime}}{\Lambda_{\chi}} \frac{C_S}{f_{\pi}^2} \epsilon_{\tau_1}(\psi) \epsilon_{\delta_3}(J/\psi) g^{\nu_1 \tau_1}\nonumber \\
&& \times \epsilon^{\lambda_2 \nu_2 \alpha_2 \beta_2} v_{\alpha_2} p_{1 \lambda_2} p_{1 \mu_2}  \epsilon^{\alpha_3 \lambda_3 \beta_3 \delta_3} v_{\alpha_3}  \Bigg\{ \frac{1}{2} \Big[ (-g_{\mu_1 \mu_2} + v_{\mu_1} v_{\mu_2}) \nonumber \\
&& \times (-g_{\nu_1 \nu_2} + v_{\nu_1} v_{\nu_2})  + (-g_{\mu_1 \nu_2} + v_{\mu_1} v_{\nu_2}) (-g_{\nu_1 \mu_2} + v_{\nu_1} v_{\mu_2}) \Big] \nonumber \\
&& - \frac{1}{3} (-g_{\mu_1 \nu_1} + v_{\mu_1} v_{\nu_1}) (-g_{\mu_2 \nu_2} + v_{\mu_2} v_{\nu_2}) \Bigg\} \nonumber \\
&& \times \frac{(-g_{\mu_1 \beta_3} + v_{\mu_1} v_{\beta_3}) (-g_{\beta_2 \lambda_3} + v_{\beta_2} v_{\lambda_3})}{[q_1^2-m_{D_s^*}^2][q_2^2-m_{D_{s2}}^2][q_3^2-m_{D^*}^2]}.
\end{eqnarray}
The velocity $v$ can be taken as (1, 0, 0, 0) in the static limit.
To model the $Z_{cs}^{(\prime)}$ resonance in the calculation of the $HH \to J/\psi K$ scattering amplitudes, a BW propagator was incorporated, which takes the form
\begin{eqnarray}
    \mbox{BW}[Z_{cs}^{(\prime)}] = (s - M^2_{Z_{cs}^{(\prime)}} + i M_{Z_{cs}^{(\prime)}} \Gamma_{Z_{cs}^{(\prime)}})^{-1},
\end{eqnarray}
with $s$ being the invariant mass of $J/\psi K$.

\section{Numerical Results}
\label{sec:results}

\subsection{Pure state analysis}

\begin{figure}[tb]
	\centering
	\includegraphics[width=0.8\linewidth]{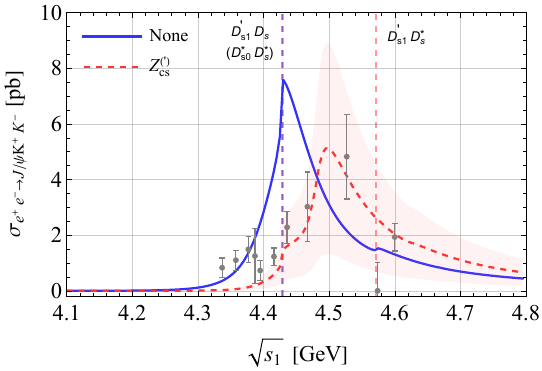}
	\caption{Cross section for $e^{+}e^{-}\to J/\psi K^{+}K^{-}$ as a function of center-of-mass energy. The solid curve represents the calculation with only the contact term included, while the dashed curve includes the $Z_{cs}^{(\prime)}$ propagator. Vertical dashed lines indicate the thresholds for $D_{s0}^* D_s^*$, $D_{s1}^\prime D_s$, and $D_{s1}^\prime D_s^*$ combinations. The shaded region represents uncertainties arising from variations in the mass and width of $Z_{cs}^{(\prime)}$.}
	\label{fig:SHH}
\end{figure}

We begin by analyzing the scenario where $\psi(4415)$ is treated as a pure $\psi(nS)$ state, utilizing the PDG average central values of $M_{\psi(4415)} = 4415$ MeV and $\Gamma_{\psi(4415)} = 110$ MeV as input parameters~\cite{ParticleDataGroup:2024cfk}. Figure~\ref{fig:SHH} presents the calculated cross section for $e^{+}e^{-}\to J/\psi K^{+}K^{-}$, incorporating contributions from both the $\psi(4415)$ resonance and the $SHH$ loops. 

When only the contact term is included (Fig.~\ref{fig:e+e-}(a)), the resulting cross section (solid curve in Fig.~\ref{fig:SHH}) is dominated by the $\psi(4415)$ resonance. This yields a pronounced peak near the $D_{s0}^* D_s^*/D_{s1}^{\prime} D_s$ threshold and a less prominent feature at the $D_{s1}^{\prime} D_s^*$ threshold. However, this prediction exhibits significant discrepancies with the experimental data, most notably the absence of the $Y(4500)$ signal around 4.5 GeV.

The inclusion of the $Z_{cs}^{(\prime)}$ state (Fig.~\ref{fig:zcs}) substantially modifies the line shape. As shown by the dashed curve in Fig.~\ref{fig:SHH}, the resulting cross section displays a prominent peak around 4.5 GeV and provides markedly improved agreement with experimental measurements. Notably, this peak position does not coincide with any conventional mass threshold in the $SHH$ loop process, suggesting its origin may be attributed to the TS mechanism arising from the rescattering processes. We therefore identify this feature as a potential explanation for the experimentally observed $Y(4500)$ state.

\begin{figure}[tb]
    \centering
	\includegraphics[width=0.8\linewidth]{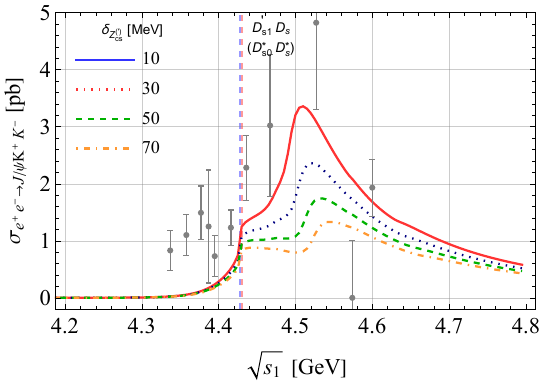}
    \caption{Dependence of the $e^{+}e^{-}\to J/\psi K^{+}K^{-}$ cross section on the mass of $Z_{cs}^{(\prime)}$, with $\delta\equiv M_{Z_{cs}^{(\prime)}}-m_{D^*}-m_{D_s^{(*)}}$ and the width fixed at $\Gamma_{Z_{cs}^{(\prime)}} = 14$ MeV. The curves demonstrate how variations in the $Z_{cs}^{(\prime)}$ mass affect the peak structure around 4.5 GeV.}
    \label{fig:mass_change}
\end{figure}

\begin{figure}[tb]
    \centering
	\includegraphics[width=0.8\linewidth]{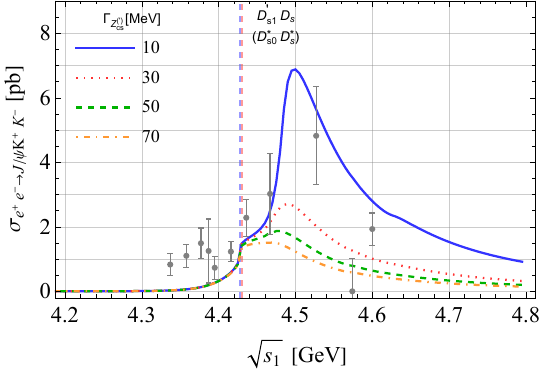}
    \caption{Dependence of the $e^{+}e^{-}\to J/\psi K^{+}K^{-}$ cross section on the width of $Z_{cs}^{(\prime)}$, with the mass fixed at $M_{Z_{cs}} = 3988$ MeV and $M_{Z_{cs}^{\prime}} = 4131$ MeV. The suppression of the 4.5 GeV peak with increasing width is clearly visible.}
    \label{fig:width_change}
\end{figure}

To quantitatively assess the sensitivity of our results to the properties of $Z_{cs}^{(\prime)}$, we systematically vary its mass and width parameters. Figures~\ref{fig:mass_change} and~\ref{fig:width_change} display the cross-section dependencies on the mass and width of $Z_{cs}^{(\prime)}$, respectively. In both cases, characteristic peak structures are observed near the $D_{s0}^* D_s^*/D_{s1}^{\prime} D_s$ threshold and around 4.5 GeV.

Figure~\ref{fig:mass_change} demonstrates that increasing the mass of $Z_{cs}^{(\prime)}$ leads to a gradual suppression and eventual disappearance of the 4.5 GeV peak. This mass dependence provides important constraints on possible $Z_{cs}^{(\prime)}$ parameters that could generate the $Y(4500)$ signal. Similarly, Fig.~\ref{fig:width_change} shows that broader $Z_{cs}^{(\prime)}$ resonances also diminish this peak structure, with the TS becoming almost completely suppressed for widths exceeding approximately 50 MeV. These systematic studies clearly demonstrate that the TS mechanism is highly sensitive to both the mass and width parameters of $Z_{cs}^{(\prime)}$.

\begin{figure}[tb]
	\centering
	\includegraphics[width=0.8\linewidth]{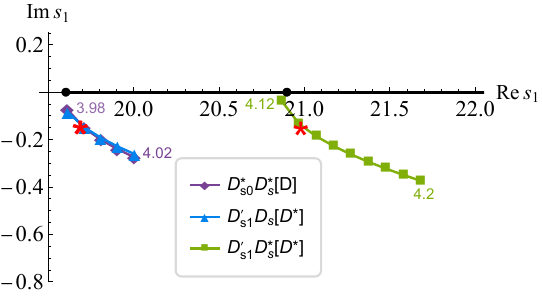}
	\caption{Trajectories of TS locations in the complex $s_1$ plane. The thick black line along the real axis represents the unitarity cut beginning at the $D_{s0}^* D_s^* (D_{s1}^{\prime} D_s)$ and $D_{s1}^{\prime} D_s^*$ thresholds. Diamond, triangle, and square symbols trace the singularity positions calculated from the $D_{s0}^* D_s^* [D]$, $D_{s1}^{\prime} D_s [D^*]$, and $D_{s1}^{\prime} D_s^* [D^*]$ loop diagrams, respectively. Annotations indicate the corresponding $Z_{cs}^{(\prime)}$ mass values in GeV. The left and right star markers highlight singularities corresponding to $M_{Z_{cs}}=3.988$ GeV and $M_{Z_{cs}^\prime}=4.131$ GeV, respectively.}
	\label{fig:s1}
\end{figure}

To further elucidate the origin of the 4.5 GeV peak, we examine the analytic structure of the scattering amplitude. The TS arises when all three intermediate particles in the loop can approach their on-shell conditions simultaneously. Namely, even in the vicinity of the physical region of the TS kinematics in the complex plane, the TS effects can manifest and significantly distort the amplitude, leading to pronounced peaks or cusps in the observed cross section~\cite{Guo:2019twa}.

The positions of these singularities can be determined by solving the Landau equation~\cite{Landau:1959fi,Guo:2019twa}. For the triangle loops considered here (Fig.~\ref{fig:e+e-}), the singularity position in $s_1$ is given by~\cite{Guo:2019twa,Liu:2015taa}
\begin{eqnarray}
s_1^- &=& (m_2 + m_3)^2 + \frac{1}{2 m_1^2} [(m_1^2 + m_2^2 - s_3) (s_2 - m_1^2 - m_3^2) \nonumber\\
&& - 4 m_1^2 m_2 m_3 - \lambda^{1/2} (s_2, m_1^2, m_3^2) \lambda^{1/2} (s_3, m_1^2, m_2^2)],
\end{eqnarray}
where $\lambda(x, y, z) = (x - y - z)^2 - 4 y z$, and $m_1$, $m_2$, $m_3$ represent the masses of the particles in the triangle loop. Here, $s_2 = (p_{K^-} + p_{J/\psi})^2$ and $s_3 = p_{K^+}^2$.
Figure~\ref{fig:s1} displays the trajectories of these singularities in the complex plane as the mass of $Z_{cs}$ varies from 3.98 to 4.02 GeV and $Z_{cs}^{\prime}$ from 4.12 to 4.2 GeV. The star markers indicate singularity positions calculated using the PDG values for $Z_{cs}$ mass and width. Crucially, these singularities lie in close proximity to the real axis, positioned to significantly influence the physical cross section in the 4.5 GeV region where the $Y(4500)$ is observed.

\subsection{S-D wave mixing scenario}

\begin{figure}[tb]
	\centering
	\includegraphics[width=0.8\linewidth]{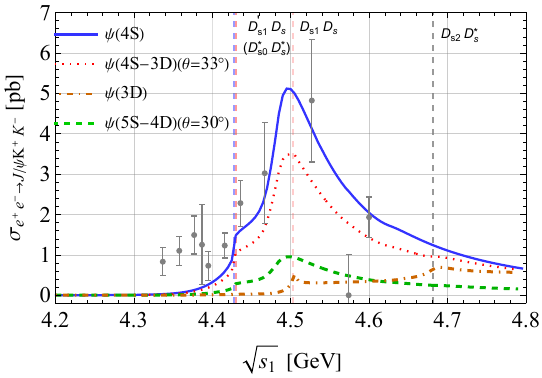}
	\caption{Cross section for $e^{+}e^{-}\to J/\psi K^{+}K^{-}$ with $\psi(4415)$ treated as a pure $S/D$-wave state or an $S-D$ mixed state. All curves include the $Z_{cs}^{(\prime)}$ propagator contribution. The solid line represents the pure $\psi(4S)$ hypothesis, the dotted line corresponds to $4S-3D$ mixing with $\theta = 33^\circ$, the dashed line shows $5S-4D$ mixing with $\theta = 30^\circ$, and the dot-dashed line represents the pure $\psi(3D)$ hypothesis.}
	\label{fig:mixing}
\end{figure}

Given the ongoing discussion in the literature regarding the exact nature of $\psi(4415)$, we also investigate scenarios where it is treated as an $S-D$ wave mixed state:
\begin{eqnarray}
\psi(4415) = |n^3S_1\rangle \cos\theta + |(n-1)^3D_1\rangle \sin\theta.
\end{eqnarray}

Figure~\ref{fig:mixing} presents cross section calculations for several specific mixing scenarios, all including the $Z_{cs}^{(\prime)}$ contribution. The solid curve represents the pure $\psi(4S)$ hypothesis, consistent with Ref.~\cite{Deng:2023mza}. The dotted curve corresponds to a $4S-3D$ mixture with a mixing angle of $33^\circ$ from Ref.~\cite{Badalian:2017nyv}, while the dashed line shows a $5S-4D$ mixture with $\theta = 30^\circ$ from Ref.~\cite{Wang:2019mhs}.

All mixing scenarios consistently produce peak structures near the $D_{s0}^* D_s^*/D_{s1}^{\prime} D_s$ threshold and around 4.5 GeV, underscoring the robustness of our interpretation regarding the $Y(4500)$ origin. A minor enhancement near the $D_{s2}D_s^*$ threshold is also observed in the mixing scenarios. To trace its origin, we computed the cross section for a pure $\psi(3D)$ state, which exhibits a distinct peak at the same location, indicating that this feature is attributed to the $D$-wave component of the charmonium wave function.

However, the cross section for $e^+e^-\to J/\psi K^+ K^-$ proceeding through the $\psi(nD)$ state and the $THH$ loop (Fig.~\ref{fig:e+e-}(b)) is significantly smaller than that via the $SHH$ loop, as shown in Fig.~\ref{fig:mixing}. This suppression can be understood from the fact that the decays $D_{s1}\to D^* K$ and $D_{s2}\to D^{(*)} K$ occur in $D$-wave in the heavy quark limit, and the corresponding rescattering amplitudes are strongly suppressed by the limited phase space. We therefore conclude that the successful description of the 4.5 GeV peak as a threshold effect in $e^+e^-\to J/\psi K^+ K^-$ within our model relies on the assumption that the $\psi(4415)$ is predominantly a $\psi(4S)$ state.

\section{Summary}
\label{sec:summary}

We have investigated the process $e^{+}e^{-} \to J/\psi K^{+}K^{-}$ via the conventional charmonia and charmed-meson rescattering loops, focusing on the origin of the $Y(4500)$ resonance observed by BESIII. In our framework, the $\psi(4415)$ state is treated as a $\psi(4S)$ pure state, as well as in $S$–$D$ mixed configurations such as $\psi(4S$–$3D)$ and $\psi(5S$–$4D)$. The role of the $Z_{cs}^{(\prime)}$ state is also examined.

When only contact interactions are considered, several cusps appear near the $SH$ thresholds. However, these contributions alone fail to reproduce the $Y(4500)$ peak near $\sqrt{s} = 4.5~\text{GeV}$. Upon introducing the $Z_{cs}^{(\prime)}$ resonance into the rescattering amplitude, a prominent peak consistent with the $Y(4500)$ signal emerges around 4.5 GeV. These results suggest that the $Y(4500)$ likely originates from the TS rather than a conventional resonance.

The interpretation remains robust regardless of whether $\psi(4415)$ is treated as a pure or mixed state. In particular, the $S$–$D$ mixing scenarios also produce the 4.5 GeV peak. Nonetheless, the dominant contribution to the $Y(4500)$-like structure comes from the $SHH$ loops with $Z_{cs}^{(\prime)}$ mediation, especially when $\psi(4415)$ is predominantly a $\psi(4S)$ state.

In conclusion, our analysis supports the interpretation of $Y(4500)$ as a threshold effect induced by the TS mechanism in the $Z_{cs}$-mediated rescattering process. Further experimental data will be essential to confirm this mechanism and clarify the nature of the states involved.

\begin{acknowledgments}
We thank the helpful discussion with Xiang Liu and Yu-Ping Guo. This work is supported by the National Natural Science Foundation of China under Grants No.~12235018, No.~11975165.

\end{acknowledgments}

\begin{appendix}
\begin{widetext}

\section{The $^{3}P_{0}$ Model}
\label{3P0 Model}

The $^{3}P_{0}$ quark pair creation model is employed to calculate the OZI-allowed strong decay processes. In this model~\cite{Micu:1968mk,LeYaouanc:1972vsx,Blundell:1995ev,Barnes:2005pb,Ackleh:1996yt}, a $q\bar{q}$ pair is created from the vacuum in a $^{3}P_{0}$ (quantum numbers of the vacuum) state. The created $q\bar{q}$ pair has orbital angular momentum $L_P = 1$, spin $S_P = 1$, and total angular momentum $J_P = 0$, so that $M_{L_P} = -M_{S_P} \equiv m$.

The transition operator for $q\bar{q}$ pair creation is given by
\begin{eqnarray}
T &=& -3\gamma \sum_{m} \langle 1 m 1 -m | 0 0 \rangle \sqrt{96 \pi} \int d^3 \vec{p}_q d^3 \vec{p}_{\bar q} {\delta}^3 (\vec{p}_q + \vec{p}_{\bar q})\nonumber\\
&& \times \mathcal{Y}_{1m} \left(\frac{\vec{p}_q - \vec{p}_{\bar q}}{2}\right) \chi_{1-m} {\phi}_0 {\omega}_0 b^{\dagger}_{q} (\vec{p}_q) d^{\dagger}_{\bar q} (\vec{p}_{\bar q}),
\end{eqnarray}
where $b^{\dagger}_{q} (\vec{p}_q)$ and $d^{\dagger}_{\bar q} (\vec{p}_{\bar q})$ denote the creation operators for the quark and antiquark, respectively. The spin triplet state of the created $q\bar{q}$ pair is described by the spin wavefunction $\chi_{1-m}$, and its momentum-space distribution is given by the solid harmonic $\mathcal{Y}_{L M_L} (\vec{k}) \equiv |\vec{k}|^L Y_{L M_L} ({\theta}_k, {\phi}_k)$. The flavor wavefunction of the created quark pair is ${\phi}_0 = \frac{1}{\sqrt{3}} (u \bar{u} + d \bar{d} + s \bar{s})$, and ${\omega}_0$ corresponds to the color singlet state. The overall factor of 3 cancels when evaluating color overlaps, and the normalization factor $\sqrt{96 \pi}$ follows conventions in Refs.~\cite{Barnes:2005pb,Ackleh:1996yt}.

The helicity amplitude for the meson decay $A \rightarrow BC$ is extracted from the transition matrix element
\begin{eqnarray}
\langle BC|T|A \rangle &=& {\delta}^3 (\vec{p}_A - \vec{p}_B - \vec{p}_C) \mathcal{M}^{M_{J_A} M_{J_B} M_{J_C}},
\end{eqnarray}
where the helicity amplitude under the normalization of Refs.~\cite{Barnes:2005pb,Ackleh:1996yt} is
\begin{eqnarray}
\mathcal{M}^{M_{J_A} M_{J_B} M_{J_C}} (\vec{P}) &=& \gamma \sum \langle L_A M_{L_A} S_A M_{S_A}|J_A M_{J_A} \rangle \langle L_B M_{L_B} S_B M_{S_B}|J_B M_{J_B} \rangle \langle L_C M_{L_C} S_C M_{S_C}|J_C M_{J_C} \rangle \nonumber\\
&& \times \langle 1 m 1 -m|00 \rangle \langle {\chi}^{14}_{S_B M_{S_B}} {\chi}^{32}_{S_C M_{S_C}}|{\chi}^{12}_{S_A M_{S_A}} {\chi}^{34}_{1 -m} \rangle \nonumber\\
&& \times \left[ \langle {\phi}_B^{14} {\phi}_C^{32}| {\phi}_A^{12} {\phi}_0^{34} \rangle I(\vec{P}, m_1, m_2, m_3) \right. \nonumber\\
&& + \left. (-1)^{1 + S_A + S_B + S_C} \langle {\phi}_B^{32} {\phi}_C^{14}| {\phi}_A^{12} {\phi}_0^{34} \rangle I(-\vec{P}, m_2, m_1, m_3) \right].
\end{eqnarray}
The two terms in the last factor correspond to the diagrams in Fig.~\ref{fig:3P0}: in diagram (a), quark 1 (antiquark 2) from meson A ends up in meson B (C), while in diagram (b), they end up in meson C (B). Here, indices 3 and 4 refer to the created quark and antiquark from the vacuum, respectively.

\begin{figure}[tb]
    \centering
	\includegraphics[width=0.4\linewidth]{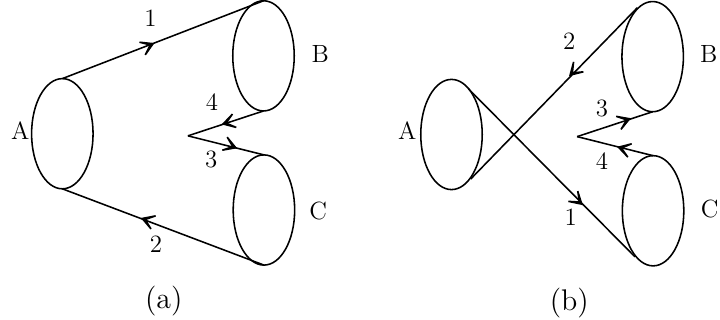}
    \caption{Two diagrams contributing to the meson decay $A \rightarrow BC$ in the $^3P_0$ model.}
    \label{fig:3P0}
\end{figure}

The momentum space integral is given by
\begin{eqnarray}
I(\vec{P}, m_1, m_2, m_3) &=& \sqrt{96 \pi} \int d^3 \vec{p} \mathcal{Y}_{lm} (\vec{p}) {\psi}_{n_A, L_A, M_{L_A}} (\vec{p} + \vec{P}) {\psi}^{\ast}_{n_B, L_B, M_{L_B}} \left(\vec{p} + \frac{m_3}{m_1 + m_3} \vec{P}\right)\nonumber\\
&& \times {\psi}^{\ast}_{n_C, L_C, M_{L_C}} \left(\vec{p} + \frac{m_3}{m_2 + m_3} \vec{P}\right),
\end{eqnarray}
where $\vec{P} \equiv \vec{P}_B = -\vec{P}_C$, and $m_1$, $m_2$, $m_3 = m_4$ are constituent quark masses. The second momentum space integral $I(-\vec{P}, m_2, m_1, m_3)$ is obtained by exchanging $B \leftrightarrow C$.

For the spatial wavefunctions, we adopt the simple harmonic oscillator (SHO) wavefunctions
\begin{eqnarray}
{\psi}_{n L M_L}(\vec{p}) = R_{nL}(p) Y_{L M_L}({\theta}_p, {\phi}_p),
\end{eqnarray}
where the radial wavefunctions are given by
\begin{eqnarray}
R_{n L}(p) &=& \frac{(-1)^n (-i)^L}{{\beta}^{\frac{3}{2}}} \sqrt{\frac{2 n!}{\Gamma(n + L + \frac{3}{2})}} \left(\frac{p}{\beta}\right)^L L_n^{L + \frac{1}{2}} \left(\frac{p^2}{{\beta}^2}\right) e^{- p^2/ (2 {\beta}^2)},
\end{eqnarray}
and $L_n^{L + \frac{1}{2}} \left(\frac{p^2}{{\beta}^2}\right)$ denotes the associated Laguerre polynomials.

The color matrix element is
\begin{eqnarray}
\langle {\omega}_B^{14} {\omega}_C^{32}| {\omega}_A^{12} {\omega}_0^{34} \rangle =
\langle {\omega}_B^{32} {\omega}_C^{14}| {\omega}_A^{12} {\omega}_0^{34} \rangle = 
\frac{1}{3}.
\end{eqnarray}

The overlap of the spin wavefunction of the meson is expressed as
\begin{eqnarray}
\langle {\chi}^{14}_{S_B M_{S_B}} {\chi}^{32}_{S_C M_{S_C}}|{\chi}^{12}_{S_A M_{S_A}} {\chi}^{34}_{1 -m} \rangle &=& (-1)^{1+ S_C} \sqrt{3(2S_A + 1) (2S_B + 1) (2S_C + 1)} \nonumber\\
&& \times \sum \langle S_B M_{S_B} S_C M_{S_C} | S M_S \rangle \langle S_A M_{S_A} 1 -m | S M_S \rangle 
\begin{Bmatrix}
\frac{1}{2} & \frac{1}{2} & S_A \\
\frac{1}{2} & \frac{1}{2} & 1 \\
S_B & S_C & S \\
\end{Bmatrix},\\
\langle {\chi}^{32}_{S_B M_{S_B}} {\chi}^{14}_{S_C M_{S_C}}|{\chi}^{12}_{S_A M_{S_A}} {\chi}^{34}_{1 -m} \rangle &=& (-1)^{1 + S_A + S_B + S_C} \langle {\chi}^{14}_{S_B M_{S_B}} {\chi}^{32}_{S_C M_{S_C}}|{\chi}^{12}_{S_A M_{S_A}} {\chi}^{34}_{1 -m} \rangle,
\end{eqnarray}
where we use the $9j$ symbol.

Using the Jacob-Wick formula~\cite{Jacob:1959at}, the helicity amplitude is converted to the partial wave amplitude $\mathcal{M}^{LS}$:
\begin{eqnarray}
\mathcal{M}^{LS} (\vec{P}) &=& \frac{\sqrt{4 \pi (2L + 1)}}{2 J_A + 1} \sum_{M_{J_B}, M_{J_C}} \langle L 0 S M_{J_A} | J_A M_{J_A} \rangle \langle J_B M_{J_B} J_C M_{J_C} | S M_{J_A} \rangle \mathcal{M}^{M_{J_A} M_{J_B} M_{J_C}} (\vec{P}).
\end{eqnarray}

The decay width is given by
\begin{eqnarray}
{\Gamma} &=& \frac{\pi P}{4 M_A^2} \sum_{LS} |\mathcal{M}^{LS}|^2,
\end{eqnarray}
where $P$ is the momentum:
\begin{eqnarray}
P &=& \frac{\sqrt{[M_A^2 - (M_B + M_C)^2] [M_A^2 - (M_B - M_C)^2]}}{2 M_A}.
\end{eqnarray}

In the numerical calculation, we set $\beta = 0.5$ GeV, $\gamma = 0.3$, and the constituent quark masses $m_{u,d} = 0.33$ GeV, $m_s = 0.55$ GeV, and $m_c = 1.5$ GeV, as in Ref.~\cite{Barnes:2005pb}.

\end{widetext}
\end{appendix}


\bibliographystyle{unsrt}

\end{document}